\documentclass{TwoCol}

\usepackage{dsfont}
\usepackage{braket}
\usepackage{slashed}
\usepackage[version=3]{mhchem}
\usepackage[colorlinks]{hyperref}
\usepackage{enumitem}
\usepackage{booktabs}

\definecolor{color1}{rgb}{0,0,0}
\definecolor{color2}{rgb}{0,0.5,0.75}

\definecolor{mylinkcolor}{rgb}{0,0,1}
\definecolor{mycitecolor}{rgb}{1,0,0}
\definecolor{myurlcolor}{rgb}{0,0.5,0}
\hypersetup{linkcolor=mylinkcolor,citecolor=mycitecolor,urlcolor=myurlcolor}

\setlist{noitemsep}

\PaperTitle{Solving the Schr\"{o}dinger Equation with Power Anharmonicity}

\Authors{Vladimir B. Belyaev\textsuperscript{1}*, Andrej Babi\v{c}\textsuperscript{2}}
\affiliation{\textsuperscript{1}\textit{Bogoliubov Laboratory of Theoretical Physics, Joint Institute for Nuclear Research, Joliot-Curie~6, 141980 Dubna, Moscow region, Russia}}
\affiliation{\textsuperscript{2}\textit{Department of Dosimetry and Application of Ionizing Radiation, Faculty of Nuclear Sciences and Physical Engineering, Czech Technical University in Prague, B\v{r}ehov\'{a}~7, 115~19 Praha~1, Czech Republic}}
\affiliation{\textbf{*E-mail:} \textit{belyaev@theor.jinr.ru}}

\Abstract{We present an application of a nonstandard approximate method---the finite-rank approximation---to solving the time-independent Schr\"{o}dinger equation for a bound-state problem. The method is illustrated on the example of a three-dimensional isotropic quantum anharmonic oscillator with additive cubic or quartic anharmonicity. Approximate energy eigenvalues are obtained and convergence of the method is discussed.}



\begin{document}
\flushbottom
\maketitle
\thispagestyle{empty}
\section{Introduction}
In molecular, atomic, nuclear and particle physics, it is often required to solve the Schr\"{o}dinger equation for a quantum system which is not exactly solvable and, in turn, it is necessary to use some approximate methods. Of special importance are the systems described by anharmonic Hamiltonians. However, in nonrelativistic quantum mechanics, apart from perturbation theory, variational method and WKB approximation, there are arguably no other standard methods applicable to these problems.

In this work, we present an application of a nonstandard method---the finite-rank approximation---to anharmonic interactions. This method has already been successfully applied to bound-state problems with both isotropic \cite{Baz61} and anisotropic \cite{Bel76} interactions, as well as low-energy scattering problems with short-range interactions \cite{Bel79}.

First, we provide a general description of the finite-rank approximation method in the context of bound-state problems in nonrelativistic quantum mechanics. Subsequently, we illustrate the method on the example of a three-dimensional isotropic quantum anharmonic oscillator with additive cubic or quartic anharmonicity. Such systems appear, e.g., in the Taylor expansion of arbitrary potential around its local minimum. Finally, we present the numerical results and discuss the convergence of the method.

\section{Finite-Rank Approximation}
\subsection{General Scheme}
We are about to solve the time-independent Schr\"{o}dinger equation for a bound-state problem:
\begin{equation}
\label{eq:tise}
H \ket{\psi} = E \ket{\psi},
\end{equation}
i.e., to find the discrete spectrum of energy eigenvalues $E_i$ and corresponding energy eigenstates $\ket{\psi_i}$ for a system defined by the Hamiltonian $H$ and certain boundary conditions.

First, let us split the full Hamiltonian $H$ into a sum of two terms:
\begin{equation}
\label{eq:h0v}
H = H_0 + V.
\end{equation}
It is convenient to choose $H_0$ so that its energy spectrum $\varepsilon_n$ and energy eigenstates $\ket{\phi_n}$ are known, satisfying the time-independent Schr\"{o}dinger equation for $H_0$:
\begin{equation}
\label{eq:tise0}
H_0 \ket{\phi_n} = \varepsilon_n \ket{\phi_n}, \qquad n = 1, \, 2, \, \dots
\end{equation}
Equation (\ref{eq:tise}) can be then written in the form:
\begin{equation}
(H_0 - E) \ket{\psi} = -V \ket{\psi}.
\end{equation}
Introducing the Green's function $G_0(E)$ of the term $H_0$:
\begin{equation}
\label{eq:g0}
G_0(E) \equiv (H_0 - E)^{-1} = \sum_n \frac{\ket{\phi_n} \bra{\phi_n}}{\varepsilon_n - E},
\end{equation}
we arrive at the Lippmann--Schwinger equation for a bound-state problem:
\begin{equation}
\label{eq:gvp}
\ket{\psi} = -G_0(E) V \ket{\psi}.
\end{equation}

This is the point where the finite-rank approximation takes place. Our ansatz will be motivated by the following operator identity \cite{Zub77}:
\begin{equation}
\label{eq:vvv}
V = V V^{-1} V = \sum_m \sum_n V \ket{\chi_m} \bra{\chi_m} V^{-1} \ket{\xi_n} \bra{\xi_n} V,
\end{equation}
where $\ket{\chi_m}$ and $\ket{\xi_n}$ form, in general, two distinct complete sets of linearly independent (not necessarily mutually orthonormal) quantum states. Indeed, in (\ref{eq:gvp}) we will approximate the term $V$, assumed to be a local operator:
\begin{equation}
\braket{\vec{r} | V | \vec{r}'} = V(\vec{r}) \, \delta^3(\vec{r} - \vec{r}'),
\end{equation}
by a nonlocal finite-rank operator $V^{(N)}$ of rank\footnote{I.e., dimension of range.} $N$, constructed from (\ref{eq:vvv}) by means of the $N$ lowest-lying energy eigenstates among $\ket{\phi_n}$ from (\ref{eq:tise0}):
\begin{equation}
\label{eq:fro}
V \approx V^{(N)} \equiv \sum_{m = 1}^N \sum_{n = 1}^N V \ket{\phi_m} (D^{-1})_{mn} \bra{\phi_n} V,
\end{equation}
where $D^{-1}$ is the inverse matrix to a matrix $D$ with elements:
\begin{equation}
D_{mn} \equiv \braket{\phi_m | V | \phi_n} = \int \mathrm{d}^3 \vec{r} \, \phi_m^*(\vec{r}) \, V(\vec{r}) \, \phi_n(\vec{r}).
\end{equation}
In turn, the only restriction in splitting the full Hamil\-ton\-ian $H$ into the sum (\ref{eq:h0v}) is the existence of the matrix $D^{-1}$.\footnote{Hence, e.g., the choice $V = 0$ is illegal.} From (\ref{eq:fro}) it follows that for $n = 1, \, 2, \, \dots \, , \, N$ the action of the operator $V^{(N)}$ on the eigenstates $\ket{\phi_n}$ is identical to that of $V$, in accordance with the interpolative character of the Bateman method \cite{Bat22}:
\begin{equation}
V^{(N)} \ket{\phi_n} = V \ket{\phi_n}, \qquad n = 1, \, 2, \, \dots \, , \, N.
\end{equation}

Substituting the operator $V^{(N)}$ into the Lippmann--Schwin\-ger equation (\ref{eq:gvp}) in place of $V$, we arrive at the ansatz:
\begin{equation}
\label{eq:oe}
\ket{\psi} = -\sum_{m = 1}^N \sum_{n = 1}^N G_0(E) V \ket{\phi_m} (D^{-1})_{mn} \braket{\phi_n | V | \psi}.
\end{equation}
Projecting onto $\bra{\phi_p} V$ and defining $b_n \equiv \braket{\phi_n | V | \psi}$, we reduce the operator---or integral---equation (\ref{eq:oe}) to a homogeneous system of $N$ linear algebraic equations:
\begin{equation}
\label{eq:ab0}
\sum_{n = 1}^N A_{pn}(E) \, b_n = 0, \qquad p = 1, \, 2, \, \dots \, , \, N
\end{equation}
for $N$ unknown amplitudes $b_n$, where the elements $A_{pn}(E)$ of a newly defined matrix $A(E)$ read:
\begin{equation}
\label{eq:a}
A_{pn}(E) \equiv \delta_{pn} + \sum_{m = 1}^N \braket{\phi_p | V G_0(E) V | \phi_m} (D^{-1})_{mn}.
\end{equation}
Nontrivial solutions of the system of equations (\ref{eq:ab0}) exist only for those values of $E$ for which the determinant of the matrix $A(E)$ vanishes:
\begin{equation}
\label{eq:deta}
\det[A(E)] = 0.
\end{equation}
The discrete set of energies $E_i$ satisfying the above equation and ordered by magnitude then constitutes the approximate spectrum of energy eigenvalues of the full Hamiltonian $H$. For each value $E_i$ the system (\ref{eq:ab0}) can be solved for $b_n$ and corresponding $i$\textsuperscript{th} approximate energy eigenstate $\ket{\psi_i}$ can be recovered from (\ref{eq:oe}) as follows:
\begin{equation}
\label{eq:psii}
\ket{\psi_i} = \sum_{m = 1}^N \sum_{n = 1}^N G_0(E_i) V \ket{\phi_m} (D^{-1})_{mn} \, b_{ni},
\end{equation}
where $b_{ni} \equiv \braket{\phi_n | V | \psi_i}$ denotes the solution $b_n$ corresponding to the energy $E_i$ and we have omitted the minus sign which does not alter the physical state. Note that the state $\ket{\psi_i}$ has to be further normalized.

\subsection{Green's Function Approximation}
If the explicit form of the Green's function $G_0(E)$ is known, one can calculate the matrix elements $\braket{\phi_p | V G_0(E) V | \phi_m}$ in (\ref{eq:a}) directly. In the position representation, we have:
\begin{align}
\braket{\phi_p | V G_0(E) V | \phi_m} = & \int \mathrm{d}^3 \vec{r} \int \mathrm{d}^3 \vec{r}' \, \phi_p^*(\vec{r}) \, V(\vec{r}) \, \times \nonumber \\
& \times \, G_0(\vec{r}, \, \vec{r}'; \, E) \, V(\vec{r}') \, \phi_m(\vec{r}').
\end{align}

On the other hand, we can proceed in our approximation even one step further and also truncate the spectral representation of the Green's function operator $G_0(E)$ from (\ref{eq:g0}) to a finite-rank operator $G_0^{(R)}(E)$ of rank $R$:
\begin{equation}
G_0(E) \approx G_0^{(R)}(E) \equiv \sum_{r = 1}^R \frac{\ket{\phi_r} \bra{\phi_r}}{\varepsilon_r - E}.
\end{equation}
This way, two free parameters---$N$ and $R$---emerge in the theory. The elements of the matrix $A(E)$ defined in (\ref{eq:a}) now take the form:
\begin{equation}
\label{eq:ar}
A_{pn}(E) = \delta_{pn} + \sum_{m = 1}^N \sum_{r = 1}^R \braket{\phi_p | V | \phi_r} \frac{1}{\varepsilon_r - E} \braket{\phi_r | V | \phi_m} (D^{-1})_{mn}
\end{equation}
and for $i$\textsuperscript{th} approximate energy eigenstate $\ket{\psi_i}$ from (\ref{eq:psii}) we get:
\begin{equation}
\label{eq:psiir}
\ket{\psi_i} = \sum_{m = 1}^N \sum_{n = 1}^N \sum_{r = 1}^R \ket{\phi_r} \frac{1}{\varepsilon_r - E_i} \braket{\phi_r | V | \phi_m} (D^{-1})_{mn} \, b_{ni}.
\end{equation}

Note that there is no point in choosing $R < N$. Indeed, if $R = c$ and $N = C$, where $c < C$, the approximate solutions $E_i$ and $\ket{\psi_i}$ reduce to those obtained when both $R = c$ and $N = c$. In the special case when we choose $R = N$, the matrix elements $A_{pn}(E)$ from (\ref{eq:ar}) further simplify to:
\begin{equation}
A_{pn}(E) = \delta_{pn} + \frac{\braket{\phi_p | V | \phi_n}}{\varepsilon_n - E},
\end{equation}
leading to results $E_i$ identical to those obtained via the standard Bubnov--Galerkin method \cite{Gal15}. Furthermore, in the relevant case $R \ge N$, the expression (\ref{eq:psiir}) for $\ket{\psi_i}$ simplifies to the following form:
\begin{equation}
\label{eq:psiirn}
\ket{\psi_i} = \sum_{n = 1}^N \ket{\phi_n} \frac{b_{ni}}{\varepsilon_n - E_i}.
\end{equation}
Projecting the result (\ref{eq:psiirn}) into the position representation, we find an expression for the $i$\textsuperscript{th} approximate energy eigenfunction $\psi_i(\vec{r})$ in terms of a finite linear combination of the first $N$ energy eigenfunctions $\phi_n(\vec{r})$:
\begin{equation}
\psi_i(\vec{r}) = \sum_{n = 1}^N \frac{b_{ni}}{\varepsilon_n - E_i} \, \phi_n(\vec{r}),
\end{equation}
also known in the literature as coupled-channel expansion.

\section{Quantum Anharmonic Oscillator}
\subsection{Full Hamiltonian}
We require the solution of the time-independent Schr\"{o}dinger equation (\ref{eq:tise}) for a three-dimensional isotropic quantum anharmonic oscillator defined by the Hamiltonian $H$ which in the position representation takes the form:
\begin{equation}
\label{eq:h}
H = \underbrace{-\frac{\hbar^2}{2 \mu} \, \nabla^2 + \frac{1}{2} \mu \omega^2 r^2}_{H_0} + \underbrace{\Lambda_P r^P}_V, \qquad P = 3, \, 4.
\end{equation}
With this choice of $H_0$ and $V$, the term $H_0$ represents the Hamiltonian of a three-dimensional isotropic quantum harmonic oscillator with intrinsic parameters $\mu$ (mass) and $\omega$ (angular frequency) and the term $V$ plays the role of an additive spherically symmetric cubic ($P = 3$) or quartic ($P = 4$) anharmonicity with coupling constant $\Lambda_P$.

\subsection{Quantum Harmonic Oscillator}
Energy eigenvalues $\varepsilon_{kl}$ and corresponding energy eigenstates $\ket{\phi_{klm}}$ of the term $H_0$ from (\ref{eq:h}), labeled by quantum numbers $k, \, l = 0, \, 1, \, 2, \, \dots$ and $m = -l, \, \dots \, , \, l$, are well known in the literature. For the energies $\varepsilon_{kl}$ we have:
\begin{equation}
\label{eq:ekl}
\varepsilon_{kl} = \frac{1}{2} \hbar \omega \, [2 \, (2k + l) + 3].
\end{equation}
Note that the energy spectrum is degenerate. The corresponding energy eigenfunctions $\phi_{klm}(\vec{r})$ take in the position representation the following form:
\begin{equation}
\phi_{klm}(\vec{r}) = r_0^{-\frac{3}{2}} \, N_{kl} \, \rho^l \, L_k^{\left( l + \frac{1}{2} \right)}(\rho^2) \, e^{-\frac{1}{2} \rho^2} \, Y_{lm}(\vartheta, \, \varphi),
\end{equation}
where $\rho \equiv r/r_0$ is a dimensionless variable, with $r_0 \equiv \sqrt{\frac{\hbar}{\mu \omega}}$ being the natural unit of length; moreover, $L_k^{\left( l + \frac{1}{2} \right)}(\rho^2)$ are the generalized Laguerre polynomials in $\rho^2$ and $Y_{lm}(\vartheta, \, \varphi)$ are the spherical harmonics which for $m = 0$ read:
\begin{equation}
Y_{l0}(\vartheta) = \sqrt{\frac{2l + 1}{4 \pi}} \, P_l(\cos\vartheta),
\end{equation}
with $P_l(\cos\vartheta)$ being the Legendre polynomials in $\cos\vartheta$. Finally, the normalization constant $N_{kl}$ can be calculated as follows:
\begin{equation}
N_{kl} = \left\lbrace \int\limits_0^{\infty} \mathrm{d} \rho \, \rho^{2l + 2} \, \left[ L_k^{\left( l + \frac{1}{2} \right)}(\rho^2) \right]^2 \, e^{-\rho^2} \right\rbrace^{-\frac{1}{2}}.
\end{equation}

For our purpose, let us label the energy eigenvalues $\varepsilon_{kl}$ and energy eigenstates $\ket{\phi_{klm}}$ by single ordinal number $n$ according to Table \ref{tab:label}. We restrict ourselves up to the 12 lowest-lying eigenstates $\ket{\phi_n}$ with energies $\varepsilon_n$ within the analysis, as well as the case $m = 0$. States $\ket{\phi_n}$ corresponding to the same row of the table are degenerate.

\begin{table}[t]
\centering
\caption{Designation of the quantum numbers $k$ and $l$ by single ordinal number $n$, assuming $m = 0$.}
\begin{tabular}{@{}rl@{\,}l@{\,}l@{}}
\toprule
$2k + l$ & \multicolumn{3}{l@{}}{$(k, \, l)_n$} \\
\midrule
$0$ & $(0, \, 0)_1$ & & \\
$1$ & $(0, \, 1)_2$ & & \\
$2$ & $(0, \, 2)_3$ & $(1, \, 0)_4$ & \\
$3$ & $(0, \, 3)_5$ & $(1, \, 1)_6$ & \\
$4$ & $(0, \, 4)_7$ & $(1, \, 2)_8$ & $(2, \, 0)_9$ \\
$5$ & $(0, \, 5)_{10}$ & $(1, \, 3)_{11}$ & $(2, \, 1)_{12}$ \\
\bottomrule
\end{tabular}
\label{tab:label}
\end{table}

\pagebreak

\section{Results \& Discussion}
\subsection{Numerical Results}
In order to measure the strength of the anharmonicity $V$ in (\ref{eq:h}), we define the dimensionless coupling constant $\lambda$ as follows:
\begin{equation}
\lambda \equiv \frac{\Lambda_P r_0^P}{\frac{1}{2} \hbar \omega}.
\end{equation}

In Tables \ref{tab:l001p3}--\ref{tab:l1p4}, we present the numerical results for the approximate energy eigenvalues $E_i$ (in the units of $\frac{1}{2} \hbar \omega$), as calculated via the finite-rank approximation method in both cases $P = 3$ and $P = 4$ while taking into account the following values of $\lambda$: $0.01$, $0.1$ and $1$. We demonstrate the convergence of the method with respect to $R$ for fixed values of $N = 4$ and $N = 8$ via comparison with the most accurate results obtained for $N = 12$.

\begin{table}[p]
\centering
\caption{$E_i \, \left[ \frac{1}{2} \hbar \omega \right]$ in the case: $\lambda = 0.01$, $P = 3$.}
\begin{tabular}{@{}rr@{.}l@{}r@{.}l@{}r@{.}l@{}r@{.}l@{}r@{.}lr@{.}l@{}}
\toprule
& \multicolumn{6}{c}{$N = 4$} & \multicolumn{4}{c}{$N = 8$} & \multicolumn{2}{c@{}}{$N = 12$} \\
\cmidrule(lr){2-7}
\cmidrule(lr){8-11}
\cmidrule(l){12-13}
$i$ & \multicolumn{2}{c}{$R = 4$} & \multicolumn{2}{c}{$R = 8$} & \multicolumn{2}{c}{$R = 12$} & \multicolumn{2}{c}{$R = 8$} & \multicolumn{2}{c}{$R = 12$} & \multicolumn{2}{c@{}}{$R = 12$} \\
\midrule
$1$ & $3$ & $022$ & $3$ & $022$ & $3$ & $022$ & $3$ & $022$ & $3$ & $022$ & $3$ & $022$ \\
$2$ & $5$ & $045$ & $5$ & $045$ & $5$ & $045$ & $5$ & $045$ & $5$ & $045$ & $5$ & $045$ \\
$3$ & $7$ & $072$ & $7$ & $071$ & $7$ & $071$ & $7$ & $071$ & $7$ & $071$ & $7$ & $071$ \\
$4$ & $7$ & $079$ & $7$ & $079$ & $7$ & $078$ & $7$ & $079$ & $7$ & $078$ & $7$ & $078$ \\
$5$ & \multicolumn{2}{c}{} & $9$ & $041$ & $9$ & $041$ & $9$ & $103$ & $9$ & $102$ & $9$ & $102$ \\
$6$ & \multicolumn{2}{c}{} & $11$ & $047$ & $11$ & $047$ & $9$ & $113$ & $9$ & $111$ & $9$ & $111$ \\
$7$ & \multicolumn{2}{c}{} & \multicolumn{2}{c}{} & $11$ & $084$ & $11$ & $138$ & $11$ & $084$ & $11$ & $138$ \\
$8$ & \multicolumn{2}{c}{} & \multicolumn{2}{c}{} & $13$ & $002$ & $11$ & $150$ & $11$ & $138$ & $11$ & $150$ \\
$9$ & \multicolumn{2}{c}{} & \multicolumn{2}{c}{} & \multicolumn{2}{c}{} & \multicolumn{2}{c}{} & $11$ & $150$ & $11$ & $156$ \\
$10$ & \multicolumn{2}{c}{} & \multicolumn{2}{c}{} & \multicolumn{2}{c}{} & \multicolumn{2}{c}{} & $13$ & $053$ & $13$ & $175$ \\
$11$ & \multicolumn{2}{c}{} & \multicolumn{2}{c}{} & \multicolumn{2}{c}{} & \multicolumn{2}{c}{} & $13$ & $096$ & $13$ & $190$ \\
$12$ & \multicolumn{2}{c}{} & \multicolumn{2}{c}{} & \multicolumn{2}{c}{} & \multicolumn{2}{c}{} & \multicolumn{2}{c}{} & $13$ & $199$ \\
\bottomrule
\end{tabular}
\label{tab:l001p3}
\end{table}

\begin{table}[p]
\centering
\caption{$E_i \, \left[ \frac{1}{2} \hbar \omega \right]$ in the case: $\lambda = 0.1$, $P = 3$.}
\begin{tabular}{@{}rr@{.}l@{}r@{.}l@{}r@{.}l@{}r@{.}l@{}r@{.}lr@{.}l@{}}
\toprule
& \multicolumn{6}{c}{$N = 4$} & \multicolumn{4}{c}{$N = 8$} & \multicolumn{2}{c@{}}{$N = 12$} \\
\cmidrule(lr){2-7}
\cmidrule(lr){8-11}
\cmidrule(l){12-13}
$i$ & \multicolumn{2}{c}{$R = 4$} & \multicolumn{2}{c}{$R = 8$} & \multicolumn{2}{c}{$R = 12$} & \multicolumn{2}{c}{$R = 8$} & \multicolumn{2}{c}{$R = 12$} & \multicolumn{2}{c@{}}{$R = 12$} \\
\midrule
$1$ & $3$ & $209$ & $3$ & $209$ & $3$ & $209$ & $3$ & $209$ & $3$ & $209$ & $3$ & $209$ \\
$2$ & $5$ & $451$ & $5$ & $406$ & $5$ & $405$ & $5$ & $412$ & $5$ & $412$ & $5$ & $412$ \\
$3$ & $7$ & $722$ & $7$ & $635$ & $7$ & $635$ & $7$ & $653$ & $7$ & $653$ & $7$ & $653$ \\
$4$ & $7$ & $807$ & $7$ & $807$ & $7$ & $694$ & $7$ & $807$ & $7$ & $694$ & $7$ & $710$ \\
$5$ & \multicolumn{2}{c}{} & $9$ & $452$ & $9$ & $450$ & $10$ & $032$ & $9$ & $885$ & $9$ & $925$ \\
$6$ & \multicolumn{2}{c}{} & $11$ & $552$ & $11$ & $552$ & $10$ & $167$ & $9$ & $964$ & $10$ & $005$ \\
$7$ & \multicolumn{2}{c}{} & \multicolumn{2}{c}{} & $11$ & $942$ & $12$ & $376$ & $11$ & $942$ & $12$ & $376$ \\
$8$ & \multicolumn{2}{c}{} & \multicolumn{2}{c}{} & $13$ & $017$ & $12$ & $565$ & $12$ & $376$ & $12$ & $565$ \\
$9$ & \multicolumn{2}{c}{} & \multicolumn{2}{c}{} & \multicolumn{2}{c}{} & \multicolumn{2}{c}{} & $12$ & $565$ & $12$ & $648$ \\
$10$ & \multicolumn{2}{c}{} & \multicolumn{2}{c}{} & \multicolumn{2}{c}{} & \multicolumn{2}{c}{} & $13$ & $662$ & $14$ & $751$ \\
$11$ & \multicolumn{2}{c}{} & \multicolumn{2}{c}{} & \multicolumn{2}{c}{} & \multicolumn{2}{c}{} & $14$ & $146$ & $14$ & $999$ \\
$12$ & \multicolumn{2}{c}{} & \multicolumn{2}{c}{} & \multicolumn{2}{c}{} & \multicolumn{2}{c}{} & \multicolumn{2}{c}{} & $15$ & $137$ \\
\bottomrule
\end{tabular}
\label{tab:l01p3}
\end{table}

\begin{table}[p]
\centering
\caption{$E_i \, \left[ \frac{1}{2} \hbar \omega \right]$ in the case: $\lambda = 1$, $P = 3$.}
\begin{tabular}{@{}rr@{.}l@{}r@{.}l@{}r@{.}l@{}r@{.}l@{}r@{.}lr@{.}l@{}}
\toprule
& \multicolumn{6}{c}{$N = 4$} & \multicolumn{4}{c}{$N = 8$} & \multicolumn{2}{c@{}}{$N = 12$} \\
\cmidrule(lr){2-7}
\cmidrule(lr){8-11}
\cmidrule(l){12-13}
$i$ & \multicolumn{2}{c}{$R = 4$} & \multicolumn{2}{c}{$R = 8$} & \multicolumn{2}{c}{$R = 12$} & \multicolumn{2}{c}{$R = 8$} & \multicolumn{2}{c}{$R = 12$} & \multicolumn{2}{c@{}}{$R = 12$} \\
\midrule
$1$ & $4$ & $521$ & $4$ & $521$ & $4$ & $397$ & $4$ & $521$ & $4$ & $397$ & $4$ & $446$ \\
$2$ & $9$ & $514$ & $6$ & $653$ & $6$ & $643$ & $8$ & $019$ & $7$ & $519$ & $7$ & $774$ \\
$3$ & $14$ & $222$ & $9$ & $098$ & $9$ & $098$ & $11$ & $846$ & $10$ & $559$ & $11$ & $846$ \\
$4$ & $15$ & $635$ & $15$ & $635$ & $10$ & $559$ & $15$ & $635$ & $11$ & $420$ & $12$ & $443$ \\
$5$ & \multicolumn{2}{c}{} & $15$ & $923$ & $12$ & $779$ & $19$ & $317$ & $11$ & $846$ & $15$ & $977$ \\
$6$ & \multicolumn{2}{c}{} & $20$ & $766$ & $16$ & $299$ & $21$ & $779$ & $13$ & $767$ & $17$ & $081$ \\
$7$ & \multicolumn{2}{c}{} & \multicolumn{2}{c}{} & $20$ & $766$ & $24$ & $756$ & $24$ & $492$ & $24$ & $756$ \\
$8$ & \multicolumn{2}{c}{} & \multicolumn{2}{c}{} & $24$ & $492$ & $28$ & $335$ & $24$ & $756$ & $28$ & $335$ \\
$9$ & \multicolumn{2}{c}{} & \multicolumn{2}{c}{} & \multicolumn{2}{c}{} & \multicolumn{2}{c}{} & $26$ & $055$ & $29$ & $781$ \\
$10$ & \multicolumn{2}{c}{} & \multicolumn{2}{c}{} & \multicolumn{2}{c}{} & \multicolumn{2}{c}{} & $28$ & $335$ & $30$ & $507$ \\
$11$ & \multicolumn{2}{c}{} & \multicolumn{2}{c}{} & \multicolumn{2}{c}{} & \multicolumn{2}{c}{} & $30$ & $942$ & $35$ & $253$ \\
$12$ & \multicolumn{2}{c}{} & \multicolumn{2}{c}{} & \multicolumn{2}{c}{} & \multicolumn{2}{c}{} & \multicolumn{2}{c}{} & $37$ & $690$ \\
\bottomrule
\end{tabular}
\label{tab:l1p3}
\end{table}

\begin{table}[p]
\centering
\caption{$E_i \, \left[ \frac{1}{2} \hbar \omega \right]$ in the case: $\lambda = 0.01$, $P = 4$.}
\begin{tabular}{@{}rr@{.}l@{}r@{.}l@{}r@{.}l@{}r@{.}l@{}r@{.}lr@{.}l@{}}
\toprule
& \multicolumn{6}{c}{$N = 4$} & \multicolumn{4}{c}{$N = 8$} & \multicolumn{2}{c@{}}{$N = 12$} \\
\cmidrule(lr){2-7}
\cmidrule(lr){8-11}
\cmidrule(l){12-13}
$i$ & \multicolumn{2}{c}{$R = 4$} & \multicolumn{2}{c}{$R = 8$} & \multicolumn{2}{c}{$R = 12$} & \multicolumn{2}{c}{$R = 8$} & \multicolumn{2}{c}{$R = 12$} & \multicolumn{2}{c@{}}{$R = 12$} \\
\midrule
$1$ & $3$ & $037$ & $3$ & $037$ & $3$ & $037$ & $3$ & $037$ & $3$ & $037$ & $3$ & $037$ \\
$2$ & $5$ & $088$ & $5$ & $084$ & $5$ & $084$ & $5$ & $085$ & $5$ & $084$ & $5$ & $084$ \\
$3$ & $7$ & $158$ & $7$ & $150$ & $7$ & $150$ & $7$ & $151$ & $7$ & $151$ & $7$ & $151$ \\
$4$ & $7$ & $188$ & $7$ & $188$ & $7$ & $179$ & $7$ & $188$ & $7$ & $179$ & $7$ & $179$ \\
$5$ & \multicolumn{2}{c}{} & $9$ & $143$ & $9$ & $142$ & $9$ & $248$ & $9$ & $234$ & $9$ & $235$ \\
$6$ & \multicolumn{2}{c}{} & $11$ & $187$ & $11$ & $187$ & $9$ & $300$ & $9$ & $280$ & $9$ & $281$ \\
$7$ & \multicolumn{2}{c}{} & \multicolumn{2}{c}{} & $11$ & $310$ & $11$ & $358$ & $11$ & $310$ & $11$ & $358$ \\
$8$ & \multicolumn{2}{c}{} & \multicolumn{2}{c}{} & $13$ & $021$ & $11$ & $434$ & $11$ & $358$ & $11$ & $434$ \\
$9$ & \multicolumn{2}{c}{} & \multicolumn{2}{c}{} & \multicolumn{2}{c}{} & \multicolumn{2}{c}{} & $11$ & $434$ & $11$ & $467$ \\
$10$ & \multicolumn{2}{c}{} & \multicolumn{2}{c}{} & \multicolumn{2}{c}{} & \multicolumn{2}{c}{} & $13$ & $234$ & $13$ & $488$ \\
$11$ & \multicolumn{2}{c}{} & \multicolumn{2}{c}{} & \multicolumn{2}{c}{} & \multicolumn{2}{c}{} & $13$ & $401$ & $13$ & $590$ \\
$12$ & \multicolumn{2}{c}{} & \multicolumn{2}{c}{} & \multicolumn{2}{c}{} & \multicolumn{2}{c}{} & \multicolumn{2}{c}{} & $13$ & $647$ \\
\bottomrule
\end{tabular}
\label{tab:l001p4}
\end{table}

\begin{table}[p]
\centering
\caption{$E_i \, \left[ \frac{1}{2} \hbar \omega \right]$ in the case: $\lambda = 0.1$, $P = 4$.}
\begin{tabular}{@{}rr@{.}l@{}r@{.}l@{}r@{.}l@{}r@{.}l@{}r@{.}lr@{.}l@{}}
\toprule
& \multicolumn{6}{c}{$N = 4$} & \multicolumn{4}{c}{$N = 8$} & \multicolumn{2}{c@{}}{$N = 12$} \\
\cmidrule(lr){2-7}
\cmidrule(lr){8-11}
\cmidrule(l){12-13}
$i$ & \multicolumn{2}{c}{$R = 4$} & \multicolumn{2}{c}{$R = 8$} & \multicolumn{2}{c}{$R = 12$} & \multicolumn{2}{c}{$R = 8$} & \multicolumn{2}{c}{$R = 12$} & \multicolumn{2}{c@{}}{$R = 12$} \\
\midrule
$1$ & $3$ & $308$ & $3$ & $308$ & $3$ & $307$ & $3$ & $308$ & $3$ & $307$ & $3$ & $307$ \\
$2$ & $5$ & $875$ & $5$ & $619$ & $5$ & $608$ & $5$ & $680$ & $5$ & $680$ & $5$ & $680$ \\
$3$ & $8$ & $575$ & $7$ & $986$ & $7$ & $986$ & $8$ & $176$ & $8$ & $176$ & $8$ & $176$ \\
$4$ & $8$ & $942$ & $8$ & $942$ & $8$ & $228$ & $8$ & $942$ & $8$ & $228$ & $8$ & $370$ \\
$5$ & \multicolumn{2}{c}{} & $10$ & $656$ & $10$ & $518$ & $11$ & $475$ & $10$ & $352$ & $10$ & $793$ \\
$6$ & \multicolumn{2}{c}{} & $13$ & $389$ & $13$ & $349$ & $12$ & $170$ & $10$ & $744$ & $11$ & $112$ \\
$7$ & \multicolumn{2}{c}{} & \multicolumn{2}{c}{} & $13$ & $389$ & $14$ & $575$ & $14$ & $575$ & $14$ & $575$ \\
$8$ & \multicolumn{2}{c}{} & \multicolumn{2}{c}{} & $14$ & $714$ & $15$ & $674$ & $14$ & $714$ & $15$ & $674$ \\
$9$ & \multicolumn{2}{c}{} & \multicolumn{2}{c}{} & \multicolumn{2}{c}{} & \multicolumn{2}{c}{} & $15$ & $674$ & $16$ & $147$ \\
$10$ & \multicolumn{2}{c}{} & \multicolumn{2}{c}{} & \multicolumn{2}{c}{} & \multicolumn{2}{c}{} & $16$ & $323$ & $17$ & $875$ \\
$11$ & \multicolumn{2}{c}{} & \multicolumn{2}{c}{} & \multicolumn{2}{c}{} & \multicolumn{2}{c}{} & $18$ & $227$ & $19$ & $457$ \\
$12$ & \multicolumn{2}{c}{} & \multicolumn{2}{c}{} & \multicolumn{2}{c}{} & \multicolumn{2}{c}{} & \multicolumn{2}{c}{} & $20$ & $334$ \\
\bottomrule
\end{tabular}
\label{tab:l01p4}
\end{table}

\begin{table}[p]
\centering
\caption{$E_i \, \left[ \frac{1}{2} \hbar \omega \right]$ in the case: $\lambda = 1$, $P = 4$.}
\begin{tabular}{@{}rr@{.}l@{}r@{.}l@{}r@{.}l@{}r@{.}l@{}r@{.}lr@{.}l@{}}
\toprule
& \multicolumn{6}{c}{$N = 4$} & \multicolumn{4}{c}{$N = 8$} & \multicolumn{2}{c@{}}{$N = 12$} \\
\cmidrule(lr){2-7}
\cmidrule(lr){8-11}
\cmidrule(l){12-13}
$i$ & \multicolumn{2}{c}{$R = 4$} & \multicolumn{2}{c}{$R = 8$} & \multicolumn{2}{c}{$R = 12$} & \multicolumn{2}{c}{$R = 8$} & \multicolumn{2}{c}{$R = 12$} & \multicolumn{2}{c@{}}{$R = 12$} \\
\midrule
$1$ & $4$ & $947$ & $4$ & $947$ & $4$ & $458$ & $4$ & $947$ & $4$ & $458$ & $4$ & $676$ \\
$2$ & $13$ & $750$ & $6$ & $380$ & $6$ & $335$ & $9$ & $554$ & $7$ & $526$ & $8$ & $604$ \\
$3$ & $22$ & $750$ & $8$ & $750$ & $8$ & $750$ & $15$ & $364$ & $11$ & $032$ & $15$ & $364$ \\
$4$ & $27$ & $553$ & $27$ & $553$ & $12$ & $517$ & $27$ & $553$ & $13$ & $469$ & $17$ & $478$ \\
$5$ & \multicolumn{2}{c}{} & $30$ & $370$ & $13$ & $469$ & $33$ & $750$ & $15$ & $364$ & $22$ & $472$ \\
$6$ & \multicolumn{2}{c}{} & $43$ & $000$ & $32$ & $898$ & $42$ & $946$ & $19$ & $672$ & $26$ & $876$ \\
$7$ & \multicolumn{2}{c}{} & \multicolumn{2}{c}{} & $43$ & $000$ & $46$ & $750$ & $46$ & $750$ & $46$ & $750$ \\
$8$ & \multicolumn{2}{c}{} & \multicolumn{2}{c}{} & $55$ & $573$ & $61$ & $136$ & $55$ & $573$ & $61$ & $136$ \\
$9$ & \multicolumn{2}{c}{} & \multicolumn{2}{c}{} & \multicolumn{2}{c}{} & \multicolumn{2}{c}{} & $57$ & $718$ & $61$ & $750$ \\
$10$ & \multicolumn{2}{c}{} & \multicolumn{2}{c}{} & \multicolumn{2}{c}{} & \multicolumn{2}{c}{} & $61$ & $136$ & $67$ & $096$ \\
$11$ & \multicolumn{2}{c}{} & \multicolumn{2}{c}{} & \multicolumn{2}{c}{} & \multicolumn{2}{c}{} & $76$ & $302$ & $82$ & $028$ \\
$12$ & \multicolumn{2}{c}{} & \multicolumn{2}{c}{} & \multicolumn{2}{c}{} & \multicolumn{2}{c}{} & \multicolumn{2}{c}{} & $92$ & $771$ \\
\bottomrule
\end{tabular}
\label{tab:l1p4}
\end{table}

\subsection{Discussion of Convergence}
Given fixed $N$, with increasing $R$ the results $E_i$ for $i \le N$ approach the accurate results ($N = 12$) rather quickly. Hence, the accuracy of the $N$ lowest energies can be considerably enhanced without $N$ taking very large values.

On the other hand, the results $E_i$ for $i > N$ turn out to be irrelevant, as they deviate from the accurate results significantly.

Finally, the finite-rank approximation is applicable even to strong couplings ($\lambda \sim 1$), although the convergence tends to be slower than in the case of small perturbations.

\nocite{*}
\bibliographystyle{ieeetr}
\bibliography{SSEPA_L}

\begin{thebibliography}{1}

\bibitem{Baz61}
N.~W. Bazley and D.~W. Fox{,} {\em Phys. Rev.}, vol.~124, no.~2, pp.~483--492,
  1961.

\bibitem{Bel76}
V.~B. Belyaev, B.~F. Irgaziev, and J.~Wrzecionko{,} {\em Yad. Fiz.}, vol.~24,
  pp.~1250--1255, 1976.

\bibitem{Bel79}
V.~B. Belyaev, M.~I. Sakvarelidze, and J.~Wrzecionko{,} {\em Phys. Lett. B},
  vol.~83, no.~1, pp.~19--21, 1979.

\bibitem{Zub77}
A.~L. Zubarev{,} {\em Theor. Math. Phys.}, vol.~30, no.~1, pp.~45--51, 1977.

\bibitem{Bat22}
H.~Bateman{,} {\em Proc. Roy. Soc. A}, vol.~100, pp.~441--449, 1922.

\bibitem{Gal15}
B.~G. Galerkin{,} {\em Vestn. Inzh. Tekhn.}, vol.~1, no.~19, pp.~897--908,
  1915.

\end{thebibliography}
\end{document}